\documentclass[]{spie}  

\usepackage{amsmath,amsfonts,amssymb}
\usepackage{graphicx}
\usepackage{indentfirst}
\usepackage{caption}
\usepackage{tikz}
\usetikzlibrary{shapes.geometric, arrows}

\usepackage{amsmath}

\usepackage{subcaption}

\usepackage[colorlinks=true, allcolors=blue]{hyperref}
\title{Design and development of the HERMES Pathfinder payloads}

\author[a, b]{R.~Campana}
\author[c, d]{Y.~Evangelista}
\author[e]{F.~Fiore}
\author[f]{A.~Guzmán}
\author[e]{G.~Baroni}
\author[c]{G.~Della~Casa}
\author[c]{G.~Dilillo}
\author[f]{P.~Hedderman}
\author[a]{E.~J.~Marchesini}
\author[g]{G.~Bertuccio}
\author[c, d]{F.~Ceraudo}
\author[h]{E.~Demenev}
\author[i]{M.~Fiorini}
\author[j]{M.~Grassi}
\author[j]{P.~Malcovati}
\author[g]{F.~Mele}
\author[k]{P.~Nogara}
\author[c]{A.~Nuti}
\author[l,m]{M.~Perri}
\author[m]{S.~Pirrotta}
\author[f]{S.~Pliego-Caballero}
\author[m]{S.~Puccetti}
\author[k]{G.~Sottile}
\author[k]{F.~Russo}
\author[e]{S.~Trevisan}

\affil[a]{INAF/OAS, Via Piero Gobetti 101, Bologna, Italy}
\affil[b]{INFN Bologna, Viale Berti Pichat 6/2, Bologna, Italy}
\affil[c]{INAF/IAPS, Via del Fosso del Cavaliere 100, Rome, Italy}
\affil[d]{INFN Roma Tor Vergata, Via della Ricerca Scientifica 1, Rome, Italy}
\affil[e]{INAF/OATS, Via Giambattista Tiepolo 11, Trieste, Italy}
\affil[f]{Universit\"at Tübingen IAAT, Sand 1, T\"ubingen, Germany}
\affil[g]{Politecnico di Milano, Via Anzani 42, Como, Italy}
\affil[h]{Fondazione Bruno Kessler, Via Sommarive 23, Povo, Trento, Italy}
\affil[i]{INAF/IASF, Via Corti 12, Milano, Italy}
\affil[j]{Universit\`a di Pavia, Via Ferrata 5, Pavia, Italy}
\affil[k]{INAF/IASF, Via Ugo La Malfa 153, Palermo, Italy}
\affil[l]{INAF/OAR, Via Frascati 33, Monte Porzio Catone, Italy}
\affil[m]{Agenzia Spaziale Italiana, Via del Politecnico snc, Rome, Italy}

\authorinfo{Further author information: (Send correspondence to R.C.)\\E-mail: \url{riccardo.campana@inaf.it}}

\begin{document} 
\maketitle

\begin{abstract}

HERMES (\emph{High Energy Rapid Modular Ensemble of Satellites}) Pathfinder mission aims to observe and localize Gamma Ray Bursts (GRBs) and other transients using a constellation of nanosatellites in low-Earth orbit (LEO). Scheduled for launch in early 2025, the 3U CubeSats will host miniaturized instruments featuring a hybrid Silicon Drift Detector (SDD) and GAGG:Ce
scintillator photodetector system, sensitive to X-rays and gamma-rays across a wide energy range. Each HERMES payload contains 120 SDD cells, each with a sensitive area of 45 mm$^2$, organized into 12 matrices, reading out 60 12.1$\times$6.94$\times$15.0 mm$^3$ GAGG:Ce
scintillators. Photons interacting with an SDD are identified as X-ray events (2--60 keV), while photons in the 20--2000 keV range absorbed by the crystals produce scintillation light, which is read by two SDDs, allowing event discrimination. The detector system, including front-end and back-end electronics, a power supply unit, a chip-scale atomic clock, and a payload data handling unit, fits within a 10$\times$10$\times$10 cm$^3$ volume, weighs 1.5~kg, and has a maximum power consumption of $\sim$2~W.. This paper outlines the development of the HERMES constellation, the design and selection of the payload detectors, and laboratory testing, presenting the results of detector calibrations and environmental tests to provide a comprehensive status update of the mission.
\end{abstract}


\section{INTRODUCTION}
\label{sec:intro}  

The HERMES Pathfinder project\footnote{\url{https://www.hermes-sp.eu}} \cite{fiore20,evangelista20,evangelista22} involves a constellation of six 3U nanosatellites equipped with X and $\gamma$-ray detectors, to monitor cosmic high energy transients, such as Gamma Ray Bursts (GRBs). The primary objective of the project is to demonstrate that precise localization of high energy cosmic transients can be achieved using miniaturized hardware, significantly reducing costs and development time compared to a conventional scientific space observatory.

Funded by the Italian Ministry for Education, University, and Research, and the Italian Space Agency, the HERMES Technological Pathfinder includes three satellites. An additional three satellites are funded by the European Union Horizon 2020 Research and Innovation Programme (Grant Agreement No. 821896, HERMES Scientific Pathfinder).

A seventh, nearly identical HERMES detector was launched on December 1, 2023, onboard the SpIRIT (\emph{Space Industry Responsive Intelligent Thermal}) mission, designed by the University of Melbourne (see Baroni et al. in these proceedings\cite{baroni24}).

This paper briefly summarize the payload requirements and design (Section~\ref{sec:req}), the payload integration and calibration (Section~\ref{sec:int}), and the current status of the payload (Section~\ref{sec:conc}).

\section{HERMES PATHFINDER PAYLOAD REQUIREMENTS AND DESIGN}
\label{sec:req}  

The scientific requirements for the HERMES Pathfinder\cite{fiore22}  necessitate ambitious payload specifications, including a broad energy range, high detection efficiency, excellent energy resolution, sub-microsecond temporal resolution, and a compact, lightweight design. The system must operate reliably in varying space environments, including different temperatures and radiation levels.

The HERMES Pathfinder mission features a hybrid detector designed to measure both X-rays and $\gamma$-rays. This  ``siswich'' design employs GAGG:Ce scintillator crystals optically linked to Silicon Drift Detectors (SDDs). The SDDs, designed by INFN-Trieste and FBK in the framework of the ReDSoX collaboration\footnote{\url{http://redsox.iasfbo.inaf.it/redsox/}}, can directly detect soft X-rays up to around 30 keV (in the so-called ``X-mode''). For higher energies, the SDDs register the light generated by $\gamma$-rays in the scintillator crystals, effectively extending the detector sensitivity into the MeV range (``S-mode'').
The discrimination between the two operating modes is achieved by the fact that a crystal is optically coupled to two independent SDD cells: a trigger on only one SDD indicates an X-mode event, while a simultaneous trigger on both SDDs indicates an S-mode event.

\begin{figure}
\centering
\includegraphics[width=\textwidth]{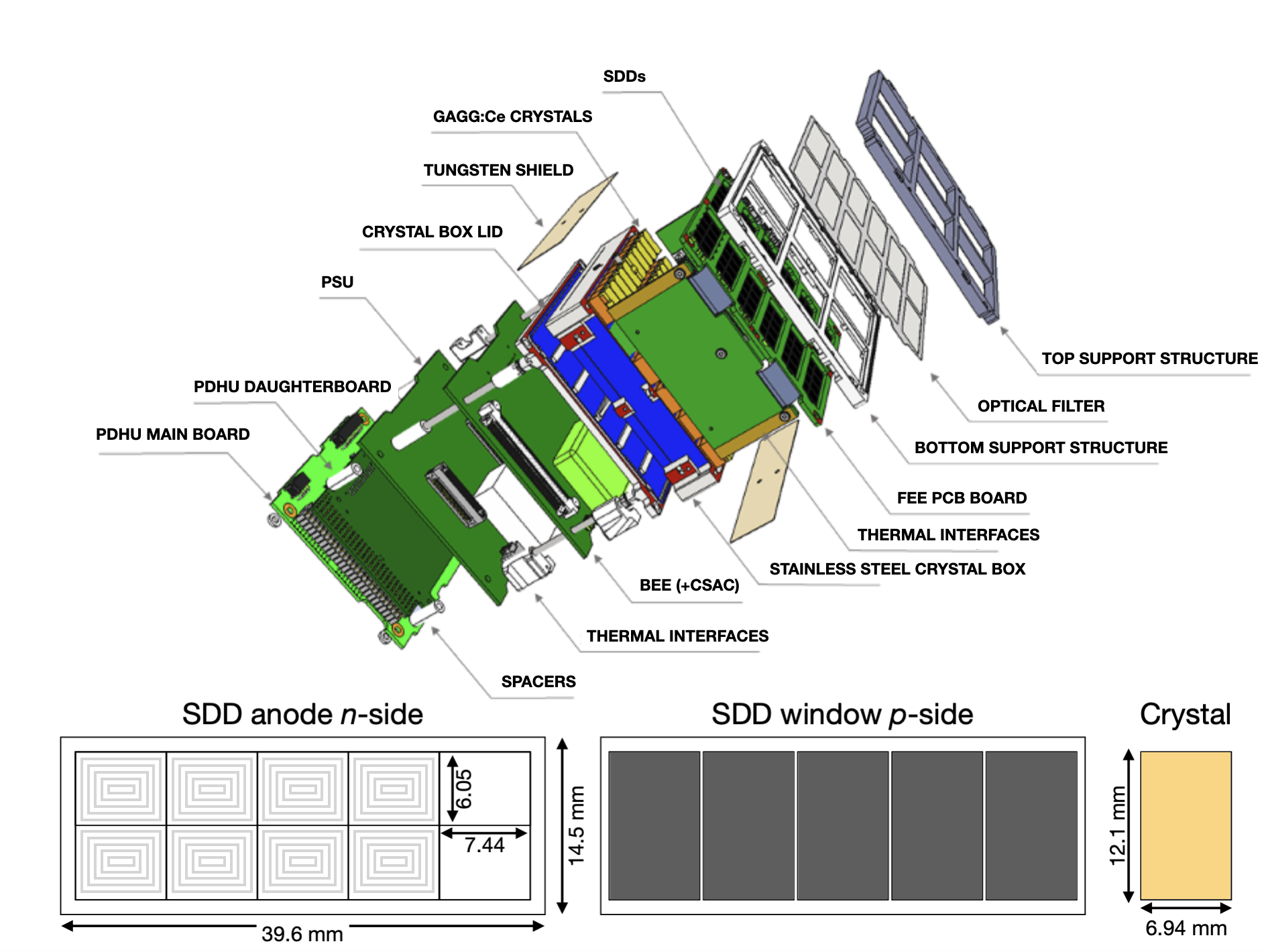}
\caption{Top panel: exploded view of the HERMES Pathfinder payload. Bottom panel: size and dimensions of the SDD matrices (both sides) and individual cells. The dimensions of the crystals (15~mm thickness) is also shown.}
\label{f:payload_exploded}
\end{figure}

Figure~\ref{f:payload_exploded} (top panel) shows an exploded view of the HERMES Pathfinder payload. 

The HERMES sensitive plane hosts a total of 120 $\sim$45 mm$^3$ SDD channels  (organized in twelve 2$\times$5 matrices, Figure~\ref{f:payload_exploded} bottom panel). Each SDD pair is coupled with a cerium-doped gadolinium-aluminium-gallium garnet (Gd$_3$Al$_2$Ga$_3$O$_{12}$:Ce or GAGG:Ce) scintillators, each 15 mm thick. The detector and crystal dimensions were chosen to maximize the effective area given all the geometrical, mechanical and physical boundary conditions imposed by the CubeSat platform and the SDD design.

The SDD detectors are hosted in a FEE board, with the readout electronics (LYRA ASIC\cite{grassi20,gandola21}) and temperature sensors. The high-performance, low-noise LYRA ASIC chipset is split in a small, single-channel die, which hosts the preamplifier and the first stage of the shaper, placed very close to the SDD anode to minimise the input stray capacitance, and a multi-channel mixed-signal logic placed on the FEE board side wings. This latter chip hosts the final shaping amplifier stages, the discriminator and the peak\&hold circuitry, with a multiplexed analog output.
The crystals are housed in a stainless-steel crystal box, with the bottom lid and two of the sides reinforced with a 200~\textmu m tungsten shielding layer. Optical coupling between crystals and SDDs is achieved by means of a $\sim$3~mm thick compressible layer of transparent silicone (Dowsil DC~93-500).
The FEE board, the upper mechanical support structures (which comprise also a thin filter to block optical light from reaching the SDDs, made by a 1~\textmu m Kapton layer coated on both sides with 150~nm Al) and the crystal box form the Detector Assembly (DA).

Below the crystal box, the back-end electronics (BEE) board hosts an Altera Cyclone V FPGA and the analog-to-digital converters. This board reads and pre-processes the analog FEE signals, managing the ASIC configuration and the event triggering. The boards houses also a chip-scale atomic clock (CSAC) which provides an extremely stable 10~MHz clock (synchronized with the GPS PPS signal) for accurate event timestamping.
The stack is then followed by a power-supply unit (PSU) boards\cite{nogara22}, which hosts the DC-DC converters which provide the necessary power supplies to the payload starting from the 3.3, 5.0 and 12.0~V lines provided by the platform, and by a Payload Data Handling Unit (PDHU\cite{guzman20}), which manages the payload and communicates with the spacecraft platform.
For redundancy, the payload is divided into four electrically and logically independent quadrants.

The payload operative modes, with transitions therein, are discussed in detail elsewhere\cite{fiore22, baroni24}. After power-on, the payload is in the \textsc{standby} mode, in which only the 3.3~V power supply line to the PDHU is switched on, and the detector is switched off. In the \textsc{ready} mode the 3.3~V and 5~V lines are turned on and the detector is operative, with the HV voltage still off. Then, in the \textsc{idle} mode the HV ramp-up is complete and the detector is fully on. Data acquisition is performed in two operative modes, \textsc{observation} (with on-board trigger search) and \textsc{calibration} (full event-by-event data recording).

\section{PAYLOAD INTEGRATION AND CALIBRATION}
\label{sec:int}  

\begin{figure}
\centering
\includegraphics[width=\textwidth]{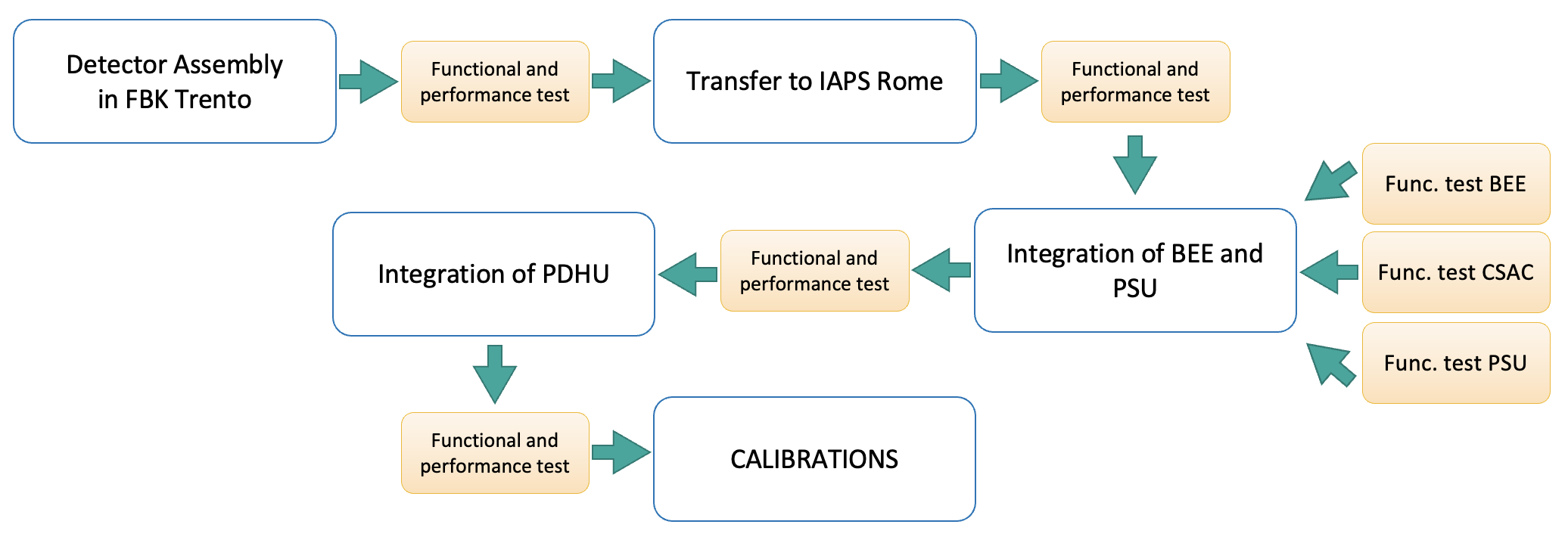}
\caption{The overall philosophy of the HERMES Pathfinder payload integration.}
\label{f:integration_flowdown}
\end{figure}

Figure~\ref{f:integration_flowdown} illustrates the overall philosophy of payload integration.

\begin{figure}
\centering
\includegraphics[width=0.5\textwidth]{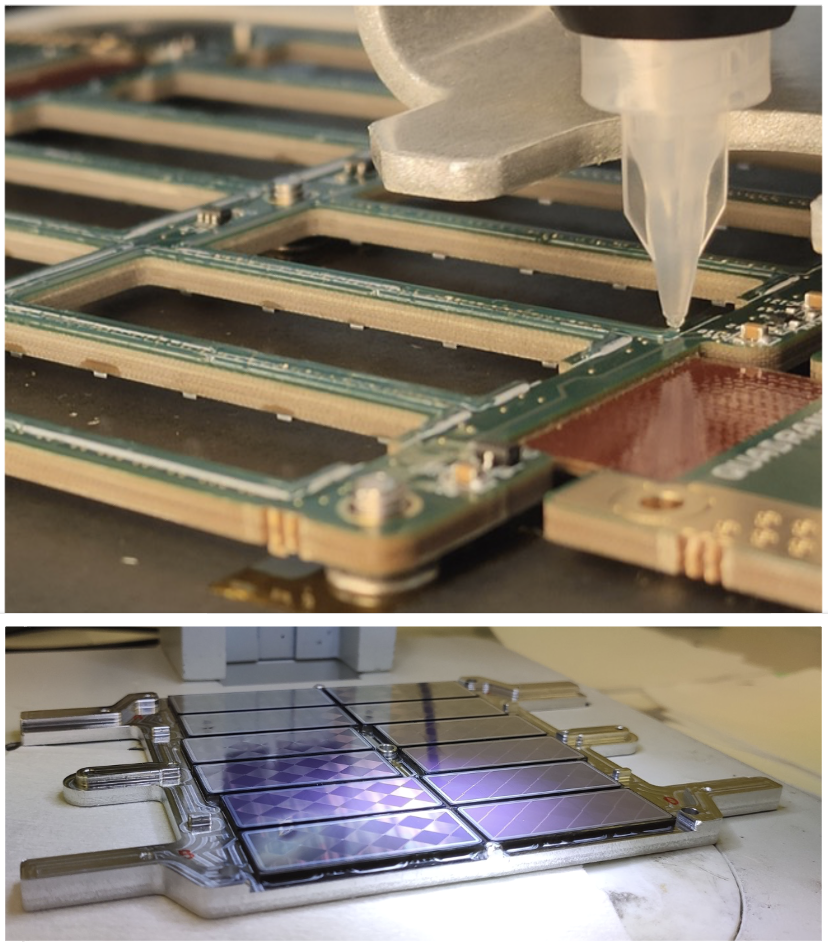}
\caption{Top panel: deposition of the glue for the SDD detectors.  Bottom panel: the SDDs integrated with the FEE board.}
\label{f:pl_integration1}
\end{figure}

The detector assembly integration occurred in the Fondazione Bruno Kessler (FBK) laboratories in Trento, Italy. FBK was also responsible for the production of the SDD sensors, which, after a quality screening, were reinforced for mechanical robustness with a Kovar frame, glued to the FEE board (Figure~\ref{f:pl_integration1}), and wire-bonded to the FEE LYRA ASICs.

\begin{figure}
\centering
\includegraphics[width=\textwidth]{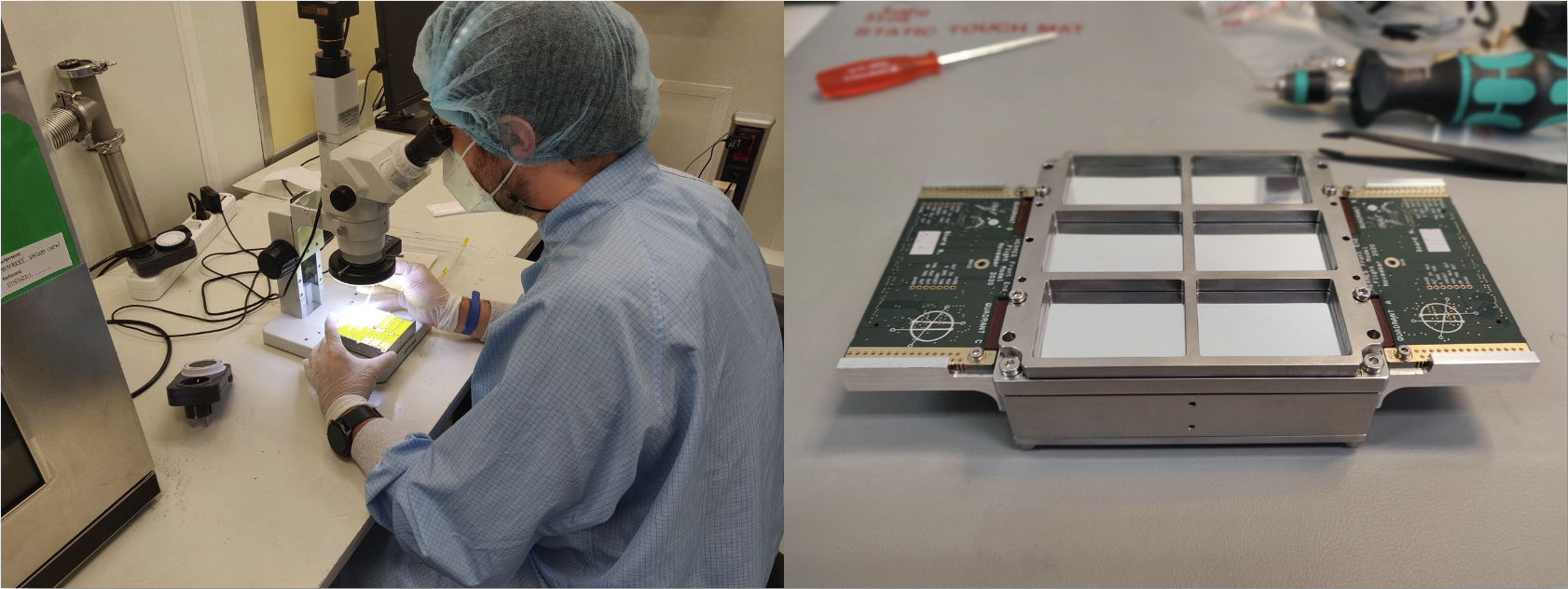}
\caption{Left panel: Optical inspection of the crystal box before integration with the FEE board.  Right panel: the complete detector assembly.}
\label{f:pl_integration2}
\end{figure}

The GAGG:Ce crystals are individually wrapped with a 3M Vikuiti Enhanced Specular Reflector (ESR), and placed in the crystal box with the soft silicone pads for optical coupling. After a careful optical inspection (Figure~\ref{f:pl_integration2}, left), the box is integrated with the FEE board (Figure~\ref{f:pl_integration2}, right).

\begin{figure}
\centering
\includegraphics[width=\textwidth]{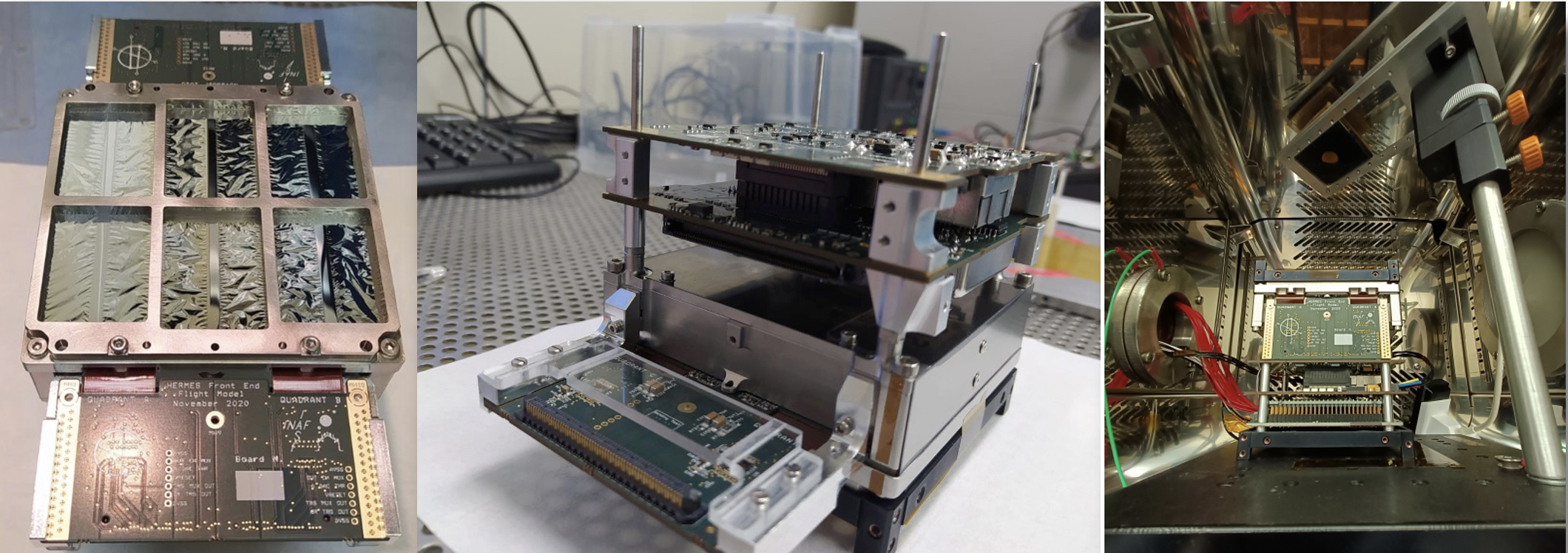}
\caption{Left: The detector assembly after integration with the optical filter. Center: The detector stack after integration of the BEE and PSU boards (upside down). Right: the final stack assembly, after PDHU integration, inside the climatic chamber ready for calibration with radioactive sources.}
\label{f:pl_integration3}
\end{figure}

Following functional and performance testing, the detector assembly is then transferred to the INAF/IAPS laboratories in Rome, where further tests were conducted at each stage of integration with at each stage of the integration with the BEE, PSU and PDHU boards (Figure~\ref{f:pl_integration3}, left and center panel).

After completing the assembly of the payload stack, the calibrations\cite{campana22,dilillo24} were performed at different temperatures (in the range $\pm$20 $^\circ$C) using different radioactive sources (e.g., $^{55}$Fe, $^{109}$Cd, $^{137}$Cs, Figure~\ref{f:pl_integration3}, right panel).

\section{CONCLUSIONS}
\label{sec:conc}  
Table~\ref{tab:masspowtab} reports the final payload mass for each of the seven flight models, and their average power consumption during operation at room temperature.

\begin{figure}
\centering
\includegraphics[width=0.4\textwidth]{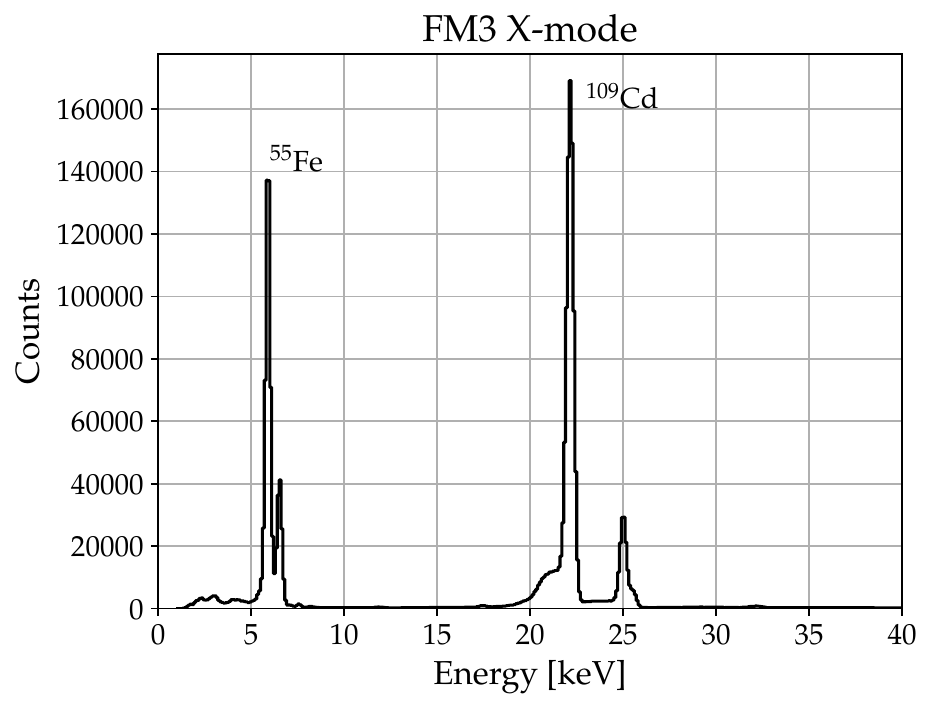}
\includegraphics[width=0.4\textwidth]{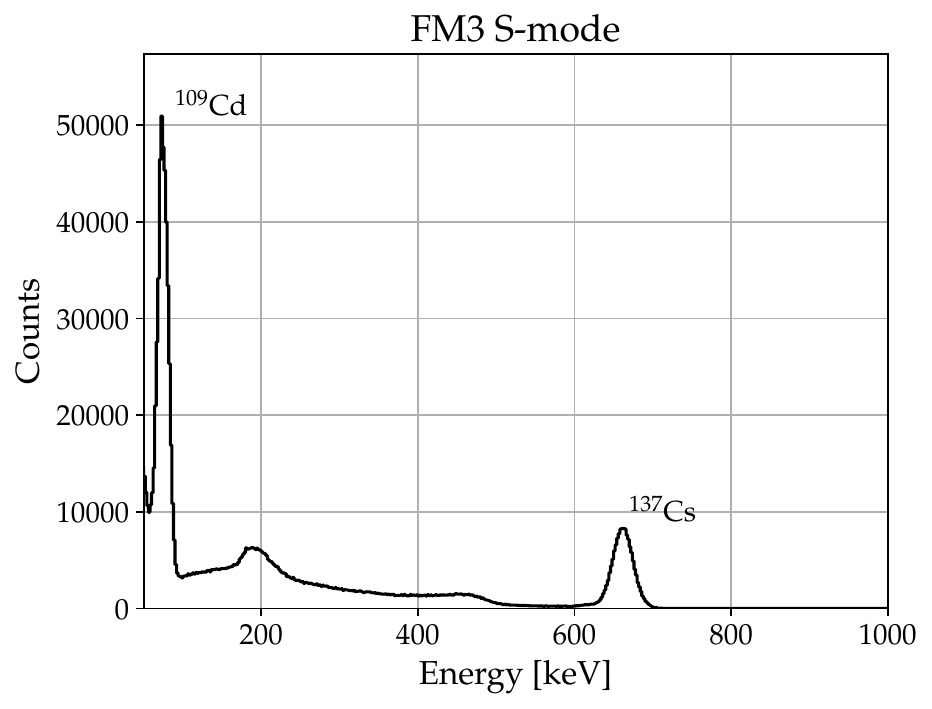}
\caption{Left panel: X-mode spectrum of $^{55}$Fe and $^{109}$Cd calibration radioactive sources. Right panel: S-mode spectrum of $^{109}$Cd and $^{137}$Cs. In this case, both spectra are acquired simultaneously with payload flight model FM3 at a temperature of 0~$^\circ$C.}
\label{f:calspectra}
\end{figure}

Figure~\ref{f:calspectra} displays a typical calibration spectrum. The average energy resolution, at an operative temperature of 0~$^\circ$C, is below 5\% at 662~keV (for S-mode events), and around 300 eV FWHM at 5.9~keV (for X-mode events). The lower threshold is $\sim$2~keV. The detector performance thus meets the requirements\cite{fiore20,fiore22,evangelista22}.

A complete discussion of payload performance after calibrations is reported in Dilillo et al. (2024, Exp. Astr., submitted).

The payload flight model FM1 has been delivered to University of Melbourne in July 2022, and integrated later that year with the SpIRIT spacecraft, which was launched December 1st, 2023. Ongoing payload commissioning\cite{baroni24}  shows expected performance from the detector in the operational scenario.
The other flight models were delivered to Politecnico di Milano (Italy) in late 2023 to mid 2024 for integration with the spacecraft. Extensive satellite-level testing (vibration and thermo-vacuum), yielded successful results. The final integration and testing phase is underway, aiming for the HERMES Pathfinder constellation launch in early 2025.

\begin{table}
\centering
\begin{tabular}{cccc}
\hline
Satellite & PL Model & Mass [g] & Power [W] \\ \hline
HERMES H1 & PFM & 1585.51 & 1.68 \\
SpIRIT & FM1 & 1562.28 & 1.63 \\
HERMES H2 & FM2 & 1573.72 & 1.63 \\
HERMES H3 & FM3 & 1581.09 & 1.69 \\
HERMES H4 & FM4 & 1575.63 & 1.67 \\
HERMES H5 & FM5 & 1592.37 & 1.68 \\
HERMES H6 & FM6 & 1579.10 & 1.68 \\ \hline
\end{tabular}
\bigskip
\caption{Payload mass and power consumption under scientific operation workload at +20~$^\circ$C. Typical measurement errors are $\pm10^{-2}$~g and $\pm10^{-2}$~W. Masses include ribs and rails for the spacecraft interface. Flight model FM1, onboard SpIRIT, has a slightly smaller mass compared to other payloads due to a different mechanical and thermal interface with the platform.} 
\label{tab:masspowtab}
\end{table}

\acknowledgments 
 
This research acknowledges support from the European Union Horizon 2020 Research and Innovation Framework Programme under grant agreements HERMES-SP n. 821896 and AHEAD2020 n. 871158, and by ASI INAF Accordo Attuativo n. 2018-10-HH.1.2020 ``HERMES--Technologic Pathfinder Attività scientifiche'' and 2022-25-HH.0 ``HERMES Pathfinder--Operazioni e sfruttamento scientifico''.

\bibliography{report} 

\begin{thebibliography}{10}

\bibitem{fiore20}
{Fiore}, F., {Burderi}, L., {Lavagna}, M., {Bertacin}, R., {Evangelista}, Y.,
  {Campana}, R., {Fuschino}, F., {Lunghi}, P., {Monge}, A., {Negri}, B.,
  {Pirrotta}, S., {Puccetti}, S., {Sanna}, A., {Amarilli}, F., {Ambrosino}, F.,
  {Amelino-Camelia}, G., {Anitra}, A., {Auricchio}, N., {Barbera}, M.,
  {Bechini}, M., {Bellutti}, P., {Bertuccio}, G., {Cao}, J., {Ceraudo}, F.,
  {Chen}, T., {Cinelli}, M., {Citossi}, M., {Clerici}, A., {Colagrossi}, A.,
  {Curzel}, S., {Della Casa}, G., {Demenev}, E., {Del Santo}, M., {Dilillo},
  G., {Di Salvo}, T., {Efremov}, P., {Feroci}, M., {Feruglio}, C., {Ferrandi},
  F., {Fiorini}, M., {Fiorito}, M., {Frontera}, F., {Gacnik}, D., {Galgoczi},
  G., {Gao}, N., {Gambino}, A.~F., {Gandola}, M., {Ghirlanda}, G., {Gomboc},
  A., {Grassi}, M., {Guzman}, A., {Karlica}, M., {Kostic}, U., {Labanti}, C.,
  {La Rosa}, G., {Lo Cicero}, U., {Lopez-Fernandez}, B., {Malcovati}, P.,
  {Maselli}, A., {Manca}, A., {Mele}, F., {Milankovich}, D., {Morgante}, G.,
  {Nava}, L., {Nogara}, P., {Ohno}, M., {Ottolina}, D., {Pasquale}, A., {Pal},
  A., {Perri}, M., {Piazzolla}, R., {Piccinin}, M., {Pliego-Caballero}, S.,
  {Prinetto}, J., {Pucacco}, G., {Rashevsky}, A., {Rashevskaya}, I., {Riggio},
  A., {Ripa}, J., {Russo}, F., {Papitto}, A., {Piranomonte}, S., {Santangelo},
  A., {Scala}, F., {Sciarrone}, G., {Selcan}, D., {Silvestrini}, S., {Sottile},
  G., {Rotovnik}, T., {Tenzer}, C., {Troisi}, I., {Vacchi}, A., {Virgilli}, E.,
  {Werner}, N., {Wang}, L., {Xu}, Y., {Zampa}, G., {Zampa}, N., and {Zanotti},
  G., ``{The HERMES-technologic and scientific pathfinder},'' in [{\em Space
  Telescopes and Instrumentation 2020: Ultraviolet to Gamma
  Ray}{\nolinebreak\hspace{0.1em}]},  {den Herder}, J.-W.~A., {Nikzad}, S., and
  {Nakazawa}, K., eds., {\em Society of Photo-Optical Instrumentation Engineers
  (SPIE) Conference Series} {\bf 11444},  114441R (Dec. 2020).

\bibitem{evangelista20}
{Evangelista}, Y., {Fiore}, F., {Fuschino}, F., {Campana}, R., {Ceraudo}, F.,
  {Demenev}, E., {Guzman}, A., {Labanti}, C., {La Rosa}, G., {Fiorini}, M.,
  {Gandola}, M., {Grassi}, M., {Mele}, F., {Morgante}, G., {Nogara}, P.,
  {Piazzolla}, R., {Pliego Caballero}, S., {Rashevskaya}, I., {Russo}, F.,
  {Sciarrone}, G., {Sottile}, G., {Milankovich}, D., {P{\'a}l}, A.,
  {Ambrosino}, F., {Auricchio}, N., {Barbera}, M., {Bellutti}, P., {Bertuccio},
  G., {Borghi}, G., {Cao}, J., {Chen}, T., {Dilillo}, G., {Feroci}, M.,
  {Ficorella}, F., {Lo Cicero}, U., {Malcovati}, P., {Morbidini}, A.,
  {Pauletta}, G., {Picciotto}, A., {Rachevski}, A., {Santangelo}, A., {Tenzer},
  C., {Vacchi}, A., {Wang}, L., {Xu}, Y., {Zampa}, G., {Zampa}, N., {Zorzi},
  N., {Burderi}, L., {Lavagna}, M., {Bertacin}, R., {Lunghi}, P., {Monge}, A.,
  {Negri}, B., {Pirrotta}, S., {Puccetti}, S., {Sanna}, A., {Amarilli}, F.,
  {Amelino-Camelia}, G., {Bechini}, M., {Citossi}, M., {Colagrossi}, A.,
  {Curzel}, S., {Della Casa}, G., {Cinelli}, M., {Del Santo}, M., {Di Salvo},
  T., {Feruglio}, C., {Ferrandi}, F., {Fiorito}, M., {Gacnik}, D.,
  {Galg{\'o}czi}, G., {Gambino}, A.~F., {Ghirlanda}, G., {Gomboc}, A.,
  {Karlica}, M., {Efremov}, P., {Kostic}, U., {Clerici}, A., {Lopez Fernandez},
  B., {Maselli}, A., {Nava}, L., {Ohno}, M., {Ottolina}, D., {Pasquale}, A.,
  {Perri}, M., {Piccinin}, M., {Prinetto}, J., {Riggio}, A., {Ripa}, J.,
  {Papitto}, A., {Piranomonte}, S., {Scala}, F., {Selcan}, D., {Silvestrini},
  S., {Rotovnik}, T., {Virgilli}, E., {Troisi}, I., {Werner}, N., {Zanotti},
  G., {Anitra}, A., {Manca}, A., and {Clerici}, A., ``{The scientific payload
  on-board the HERMES-TP and HERMES-SP CubeSat missions},'' in [{\em Space
  Telescopes and Instrumentation 2020: Ultraviolet to Gamma
  Ray}{\nolinebreak\hspace{0.1em}]},  {den Herder}, J.-W.~A., {Nikzad}, S., and
  {Nakazawa}, K., eds., {\em Society of Photo-Optical Instrumentation Engineers
  (SPIE) Conference Series} {\bf 11444},  114441T (Dec. 2020).

\bibitem{evangelista22}
{Evangelista}, Y., {Fiore}, F., {Campana}, R., {Ceraudo}, F., {Della Casa}, G.,
  {Demenev}, E., {Dilillo}, G., {Fiorini}, M., {Grassi}, M., {Guzman}, A.,
  {Hedderman}, P., {Marchesini}, E.~J., {Morgante}, G., {Mele}, F., {Nogara},
  P., {Nuti}, A., {Piazzolla}, R., {Pliego Caballero}, S., {Rashevskaya}, I.,
  {Russo}, F., {Sottile}, G., {Labanti}, C., {Baroni}, G., {Bellutti}, P.,
  {Bertuccio}, G., {Cao}, J., {Chen}, T., {Dedolli}, I., {Feroci}, M.,
  {Fuschino}, F., {Gandola}, M., {Gao}, N., {Ficorella}, F., {Malcovati}, P.,
  {Picciotto}, A., {Rachevski}, A., {Santangelo}, A., {Tenzer}, C., {Vacchi},
  A., {Wang}, L., {Xu}, Y., {Zampa}, G., {Zampa}, N., and {Zorzi}, N.,
  ``{Design, integration, and test of the scientific payloads on-board the
  HERMES constellation and the SpIRIT mission},'' in [{\em Space Telescopes and
  Instrumentation 2022: Ultraviolet to Gamma Ray}{\nolinebreak\hspace{0.1em}]},
   {den Herder}, J.-W.~A., {Nikzad}, S., and {Nakazawa}, K., eds., {\em Society
  of Photo-Optical Instrumentation Engineers (SPIE) Conference Series} {\bf
  12181},  121811G (Aug. 2022).

\bibitem{baroni24}
Baroni, G., ``{The commissioning and early operations of the high energy HERMES
  payload onboard SpIRIT},'' in [{\em Space Telescopes and Instrumentation
  2024: Ultraviolet to Gamma Ray}{\nolinebreak\hspace{0.1em}]},  {den Herder},
  J.-W.~A., {Nikzad}, S., and {Nakazawa}, K., eds., {\em Society of
  Photo-Optical Instrumentation Engineers (SPIE) Conference Series} {\bf this
  volume} (Aug. 2024).

\bibitem{fiore22}
{Fiore}, F., {Guzman}, A., {Campana}, R., and {Evangelista}, Y.,
  ``{HERMES-Pathfinder},'' in [{\em Handbook of X-ray and Gamma-ray
  Astrophysics}{\nolinebreak\hspace{0.1em}]},   38 (2022).

\bibitem{grassi20}
Grassi, M., Gandola, M., Mele, F., Bertuccio, G., Malcovati, P., Fuschino, F.,
  Campana, R., Labanti, C., Fiorini, M., Evangelista, Y., Piazzolla, R.,
  Feroci, M., Zampa, G., Zampa, N., Rachevski, A., Bellutti, P., Borghi, G.,
  Demenev, E., Ficorella, F., Picciotto, A., Zorzi, N., Rashevskaya, I.,
  Vacchi, A., Fiore, F., and Burderi, L., ``{X-/$\gamma$-Ray Detection
  Instrument for the HERMES Nano-Satellites Based on SDDs Read-Out by the LYRA
  Mixed-Signal ASIC Chipset},'' in [{\em 2020 IEEE International
  Instrumentation and Measurement Technology Conference
  (I2MTC)}{\nolinebreak\hspace{0.1em}]},   1--6 (2020).

\bibitem{gandola21}
{Gandola}, M., {Mele}, F., {Grassi}, M., {Malcovati}, P., and {Bertuccio}, G.,
  ``{Multi-chip front-end electronics LYRA for X and {\ensuremath{\gamma}} Ray
  detector for HERMES mission},'' {\em Journal of Instrumentation}~{\bf 16},
  T12013 (Dec. 2021).

\bibitem{nogara22}
{Nogara}, P., {Sottile}, G., {Russo}, F., {La Rosa}, G., {Lo Gerfo}, F.~P.,
  {Del Santo}, M., {Evangelista}, Y., {Campana}, R., {Fuschino}, F., and
  {Fiore}, F., ``{The power supply unit onboard the HERMES nano-satellite
  constellation},'' in [{\em Space Telescopes and Instrumentation 2022:
  Ultraviolet to Gamma Ray}{\nolinebreak\hspace{0.1em}]},  {den Herder},
  J.-W.~A., {Nikzad}, S., and {Nakazawa}, K., eds., {\em Society of
  Photo-Optical Instrumentation Engineers (SPIE) Conference Series} {\bf
  12181},  121815M (Aug. 2022).

\bibitem{guzman20}
{Guzman}, A., {Pliego}, S., {Bayer}, J., {Evangelista}, Y., {La Rosa}, G.,
  {Sottile}, G., {Curzel}, S., {Campana}, R., {Fiore}, F., {Fuschino}, F.,
  {Colagrossi}, A., {Fiorito}, M., {Nogara}, P., {Piazzolla}, R., {Russo}, F.,
  {Santangelo}, A., and {Tenzer}, C., ``{The Payload Data Handling Unit (PDHU)
  on-board the HERMES-TP and HERMES-SP CubeSat Missions},'' in [{\em Space
  Telescopes and Instrumentation 2020: Ultraviolet to Gamma
  Ray}{\nolinebreak\hspace{0.1em}]},  {den Herder}, J.-W.~A., {Nikzad}, S., and
  {Nakazawa}, K., eds., {\em Society of Photo-Optical Instrumentation Engineers
  (SPIE) Conference Series} {\bf 11444},  1144450 (Dec. 2020).

\bibitem{campana22}
{Campana}, R., {Baroni}, G., {Della Casa}, G., {Dilillo}, G., {Marchesini},
  E.~J., {Ceraudo}, F., {Guzm{\'a}n}, A., {Hedderman}, P., and {Evangelista},
  Y., ``{Calibration of the first detector flight models for the HERMES
  constellation and the SpIRIT mission},'' in [{\em Space Telescopes and
  Instrumentation 2022: Ultraviolet to Gamma Ray}{\nolinebreak\hspace{0.1em}]},
   {den Herder}, J.-W.~A., {Nikzad}, S., and {Nakazawa}, K., eds., {\em Society
  of Photo-Optical Instrumentation Engineers (SPIE) Conference Series} {\bf
  12181},  121815K (Aug. 2022).

\bibitem{dilillo24}
{Dilillo}, G., {Marchesini}, E.~J., {Della Casa}, G., {Baroni}, G., {Campana},
  R., {Borciani}, E., {Srivastava}, S., {Trevisan}, S., {Ceraudo}, F.,
  {Citossi}, M., {Evangelista}, Y., {Guzm{\'a}n}, A., {Hedderman}, P.,
  {Labanti}, C., {Virgilli}, E., and {Fiore}, F., ``{The HERMES calibration
  pipeline: MESCAL},'' {\em Astronomy and Computing}~{\bf 46},  100797 (Jan.
  2024).

\end{thebibliography}
\bibliographystyle{spiebib} 

\end{document}